\documentclass[]{IEEEtran} 

\usepackage{cite}
\usepackage{hyperref}

\usepackage{tikz}
\usepackage{pgfplots}
\usetikzlibrary{calc} 

\tikzstyle{block}=[draw, rectangle, text centered, minimum width=2em, minimum height=3em]
\tikzstyle{sum}=[draw, circle, minimum size=.2cm]

\usepackage{amssymb}
\usepackage{amsmath}

\begin{document}

\title{Reducing MIMO Detection Complexity\\via Hierarchical Modulation}
\author{Yi{\u{g}}it U{\u{g}}ur, \IEEEmembership{Student Member, IEEE,} and A. {\"O}zg{\"u}r Y{\i}lmaz, \IEEEmembership{Member, IEEE} 
\thanks{Y. Ugur was with the Electrical and Electronics Engineering Department, Middle East Technical University (METU), Ankara, Turkey. He is now with the Chair of Digital Communication Systems, Ruhr-Universit{\"a}t Bochum (RUB), Bochum, Germany (e-mail: yigit.ugur@rub.de).}
\thanks{A. O. Yilmaz is with the Electrical and Electronics Engineering Department, Middle East Technical University (METU), Ankara, Turkey (e-mail: aoyilmaz@metu.edu.tr).} 
}

\markboth{SUBMITTED TO IEEE COMMUNICATIONS LETTERS}{Ugur and Yilmaz: Reducing MIMO Detection Complexity via Hierarchical Modulation} 


\maketitle
\begin{abstract}
This work considers multiple-input multiple-output (MIMO) communication systems using hierarchical modulation. A disadvantage of the maximum-likelihood (ML) MIMO detector is that computational complexity increases exponentially with the number of transmit antennas. To reduce complexity, we propose a hierarchical modulation scheme to be used in MIMO transmission where base and enhancement layers are incorporated. In the proposed receiver, the base layer is detected first with a minimum mean square error (MMSE) detector which is followed by ML detection of the enhancement layer. Our results indicate that the proposed low complexity scheme does not compromise performance when design parameters such as code rates and constellation ratio are chosen carefully.
\end{abstract}


\section{Introduction}
\label{sec:Introduction}

\IEEEPARstart{M}{IMO} (multiple-input multiple-output) communication systems can provide high data rates through spatial multiplexing, in which independent data streams are sent from different antennas. The optimal maximum-likelihood (ML) detection method for MIMO systems has high computational complexity, whose order is exponential with the number of transmit antennas. Complexity can be reduced by linear receivers such as zero forcing (ZF) and linear minimum mean square error (MMSE) receivers. There are numerous proposals that provide a reasonable trade-off between complexity and performance. The sphere decoding (SD) algorithm~\cite{fincke}, the ML detection with QR Decomposition and M-algorithm (QRM-MLD)~\cite{kawai} and graph based detection based on the belief propagation (BP) algorihm~\cite{BP} are some of the commonly used methods to decrease complexity. A different approach is transforming the MIMO ML detection problem to a convex optimization problem with semidefinite programming (SDP) relaxation~\cite{wiesel}. Hieararchical modulation was proposed by~\cite{cover} and has been long in use for various purposes, yet its utilization for receiver complexity reduction is novel to the best of our knowledge and stands as the main contribution of this paper.   

In this work, a high data rate MIMO system with low receiver computational complexity is developed. Considering ML and MMSE receivers, a two-stage receiver structure is utilized. Multiple layers are transmitted using hierarchical modulation, where parameters such as code rates at each layer are adapted according to ML and MMSE receivers' error rate performance capabilities. A similar idea is studied in~\cite{rupp}, where ML is used at both stages of the receiver and bit error rate (BER) performance exhibits an error floor. In our proposed receiver, the base layer is detected first with the MMSE filter which is followed by ML detection of the enhancement layer. The proposed receiver has a better error rate performance compared to~\cite{rupp}, since it takes into account distinct capabilities of ML and MMSE receivers, which will be explained later in Section~\ref{sec:WhyOrder}. Our structure is hardware friendly to implement in that computational complexity just depends on the number of antennas and the size of constellation sent in one layer.   

\IEEEpubidadjcol 

The remainder of this paper is organized as follows. The system model is described in Section~\ref{sec:SystemModel}. The proposed two-stage receiver structure is described in Section~\ref{sec:ProposedReciver}. Simulation results are presented in Section~\ref{sec:SimResults}. Finally, conclusions are drawn in Section~\ref{sec:Conclusion}.

\vspace{0.8em}
\emph{Notation:} Bold small and capital letters denote vector and matrices, respectively. The superscripts $(\cdot)^{\text{H}}$, $(\cdot)^{{\text{T}}}$ and $(\cdot)^{-1}$ represent the Hermitian transpose, the transpose and the matrix inverse, respectively. The identity matrix is represented by $\mathbf{I}$. The 
expectation is denoted by $\mathrm{E}[\cdot]$. Big-O Notation is denoted by $\mathcal{O}(\cdot)$ and trace operation is represented by $\mathrm{tr}(\cdot)$. $\mathbb{C}^{x\times y}$ denotes the space of $x\times y$ complex matrices. The floor function is denoted by $\lfloor \cdot \rfloor$, where  $\lfloor x \rfloor$ represents the largest integer smaller than or equal to $x$. 

\section{System Model}
\label{sec:SystemModel}
Gray-mapped 16-ary hierarchical quadrature amplitude modulation (16-HQAM) is used in this work to enable sequential detection of sub-constellations (layers). The average power of constellation points is set to unity without loss of generality. The minimum distance of the base layer constellation points is represented by $d_{1}$ and the minimum distance of the enhancement layer constellation points is represented by $d_{2}$. The ratio $d=d_{1}/d_{2}$ is called the constellation ratio. For the case $d=2$, the constellation corresponds to that of standard \mbox{16-QAM}. Protection levels of layers can be arranged by changing $d$. When $d$ increases, the base layer has better error rate performance meanwhile enhancement layer has worse.    

\begin{figure*}[t!]
\centering	
\begin{tikzpicture}[thick,scale=1]
\node (in1) at (0,0) {};
\node (in2) at (0,3) {};
\node (enc1) at (1.5,0) [block, align=center] {Encoder for\\Enh. Layer};
\node (enc2) at (1.5,3) [block, align=center] {Encoder for\\Base Layer};
\draw[->] (in1) to (enc1);
\draw[->] (in2) to (enc2);
\node (mod) at (3.5,1.5) [block, align=center] {Modulator};
\draw[->] (enc1) -| (mod);
\draw[->] (enc2) -| (mod);
\node (demux) at (5.5,1.5) [block, align=center, minimum height=3.8cm] {\rotatebox{90}{Demultiplexer}};	
\draw[->] (mod) to (demux);
	
\coordinate (ant2_lb) at ($(demux.north east)-(0,0.5)$);
\coordinate (ant2_lf) at ($(demux.north east)-(0,0.5)+(0.5,0)$);
\draw (ant2_lb) to (ant2_lf) node[below] {$x_{1}$};
	
\coordinate (ant1_lb) at ($(demux.south east)+(0,0.5)$);
\coordinate (ant1_lf) at ($(demux.south east)+(0,0.5)+(0.5,0)$);
\draw (ant1_lb) to (ant1_lf) node[below] {$x_{N_{t}}$};
	
\draw[fill=black] ($(ant1_lf.center)+(0,1)$) circle (0.04cm);
\draw[fill=black] ($(ant1_lf.center)+(0,1.4)$) circle (0.04cm);
\draw[fill=black] ($(ant1_lf.center)+(0,1.8)$) circle (0.04cm);
\draw[fill=black] ($(ant1_lf.center)+(0,2.2)$) circle (0.04cm);
		
\draw[scale=0.1] ($(ant2_lf.center)+(0,3)$) -- ($(ant2_lf.center)+(1.5,3)+sqrt(3)*(0,1.5)$) -- ($(ant2_lf.center)+(-1.5,3)+sqrt(3)*(0,1.5)$) -- cycle;
\draw[scale=0.1] (ant2_lf.center) -- ($(ant2_lf)+(0,3)+sqrt(3)*(0,1.5)$);
	
\draw[scale=0.1] ($(ant1_lf.center)+(0,3)$) -- ($(ant1_lf.center)+(1.5,3)+sqrt(3)*(0,1.5)$) --  ($(ant1_lf.center)+(-1.5,3)+sqrt(3)*(0,1.5)$) -- cycle;
\draw[scale=0.1] (ant1_lf.center) -- ($(ant1_lf)+(0,3)+sqrt(3)*(0,1.5)$);
	
\coordinate (ant1_r) at ($(demux.south east)+(1.3,0.5)$);
\coordinate (ant2_r) at ($(demux.north east)+(1.3,-0.5)$);
	
\draw[fill=black] ($(ant1_r.center)+(0,1)$) circle (0.04cm);
\draw[fill=black] ($(ant1_r.center)+(0,1.4)$) circle (0.04cm);
\draw[fill=black] ($(ant1_r.center)+(0,1.8)$) circle (0.04cm);
\draw[fill=black] ($(ant1_r.center)+(0,2.2)$) circle (0.04cm);   
	
\draw[scale=0.1] ($(ant2_r.center)+(0,3)$) -- ($(ant2_r.center)+(1.5,3)+sqrt(3)*(0,1.5)$) --  ($(ant2_r.center)+(-1.5,3)+sqrt(3)*(0,1.5)$) -- cycle;
\draw[scale=0.1] (ant2_r.center) -- ($(ant2_r)+(0,3)+sqrt(3)*(0,1.5)$);
	
\draw[scale=0.1] ($(ant1_r.center)+(0,3)$) -- ($(ant1_r.center)+(1.5,3)+sqrt(3)*(0,1.5)$) --  ($(ant1_r.center)+(-1.5,3)+sqrt(3)*(0,1.5)$) -- cycle;
\draw[scale=0.1] (ant1_r.center) -- ($(ant1_r)+(0,3)+sqrt(3)*(0,1.5)$);  	
	
\node (sum1) at ($(ant1_r)+(0.7,0)$) [sum] {};	
\draw ($(sum1.center)+(-0.1,0)$) to ($(sum1.center)+(0.1,0)$);
\draw ($(sum1.center)+(0,-0.1)$) to ($(sum1.center)+(0,0.1)$); 
	
\node (sum2) at ($(ant2_r)+(0.7,0)$) [sum] {};	
\draw ($(sum2.center)+(-0.1,0)$) to ($(sum2.center)+(0.1,0)$);
\draw ($(sum2.center)+(0,-0.1)$) to ($(sum2.center)+(0,0.1)$);
	
\draw (ant1_r) to (sum1.west);
\draw (ant2_r) to (sum2.west);
	
\node (n1) at ($(sum1.north)+(0,0.5)$) {$n_{N_{r}}$};	
\draw[->] (n1) to (sum1.north);
	
\node (n2) at ($(sum2.north)+(0,0.5)$) {$n_{1}$};	
\draw[->] (n2) to (sum2.north);	
	
\coordinate (begin1) at (8.5,1.5) {};	
\draw (sum1.east) -| node[below] {$y_{N_{r}}$} (begin1.center);
\draw (sum2.east) -| node[above] {$y_{1}$}	 (begin1.center);
\coordinate (begin2) at (9,1.5) {};
\draw (begin1) to (begin2);			
	
\node (cancel) at ($(begin2)+(1.3,-1.5)$) [block, align=center] {Base Layer\\Cancelation};
\node (mmse) at ($(begin2)+(1.3,1.5)$) [block, align=center, minimum width=1.9cm] {MMSE\\Receiver};
\draw[->] (begin2.center) |- (cancel.west);
\draw[->] (begin2.center) |- (mmse.west);
	
\node (llr1) at ($(cancel)+(3,0)$) [block, align=center] {LLR Calculation\\(ML) $\&$ Multiplexer};
\node (llr2) at ($(mmse)+(3,0)$) [block, align=center, minimum width=3.1cm] {LLR Calculation\\$\&$ Multiplexer};
\draw[->] (cancel) to (llr1);
\draw[->] (mmse) to (llr2);
	
\node (dec1) at ($(llr1)+(3,0)$) [block, align=center] {Decoder for\\Enh. Layer};
\node (dec2) at ($(llr2)+(3,0)$) [block, align=center] {Decoder for\\Base Layer};
\draw[->] (llr1) to (dec1);
\draw[->] (llr2) to (dec2);

\node (fin1) at ($(dec1.east)+(0.8,0)$) {};
\node (fin1_r) at ($(dec1.east)+(0.45,0.3)$) {$\hat{\mathbf{x}}_{e}$};
\node (fin2) at ($(dec2.east)+(0.8,0)$) {};
\node (fin2_r) at ($(dec2.east)+(0.45,0.3)$) {$\hat{\mathbf{x}}_{b}$};
\draw[->] (dec1) to (fin1);
\draw[->] (dec2) to (fin2);

\coordinate (cizgib) at ($(dec2.east)+(0.1,0)$);
\coordinate (cizgif) at ($(dec2.east)+(0.1,-1.5)$);
\draw (cizgib) to (cizgif);
\draw[->] (cizgif) -| (cancel.north);
	
\end{tikzpicture}
\caption{System structure of our proposed receiver.}
\label{fig:Structure}
\end{figure*}
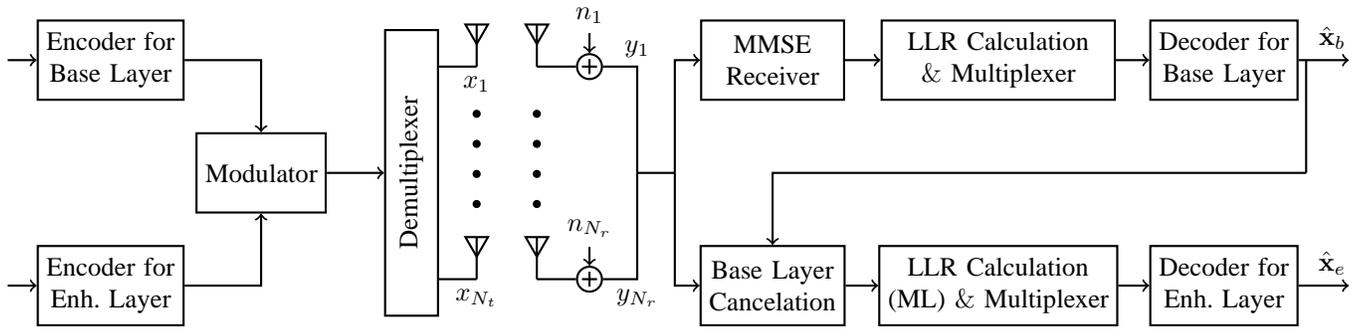
	
We consider a single point-to-point $N_{t} \times N_{r}$ MIMO system, where $N_{t}$ and $N_{r}$ are the number of transmit and receive antennas, respectively ($N_{t} \leq N_{r}$). The received signal vector \mbox{$\mathbf{y} \in \mathbb{C}^{N_{r}\times1}$}  is given by     	
\begin{align}
\mathbf{y}=\frac{1}{\sqrt{N_{t}}}\mathbf{Hx} + \mathbf{n} ,
\label{eq:basicMIMO}
\end{align}
where \mbox{$\mathbf{x}=[x_{1}, ... , x_{N_{t}}]^{\text{T}} \in \mathbb{C}^{N_{t}\times1}$} is the transmitted signal vector with $\mathrm{E}[\mathbf{x}\mathbf{x}^{\text{H}}]=\mathbf{I}$, $\mathbf{n} \in \mathbb{C}^{N_{r}\times1}$ is the independent and identically distributed, zero-mean, proper additive white Gaussian noise vector with variance $N_{0}$ and \mbox{$\mathbf{H} = [\mathbf{h}_{1}, ... ,\mathbf{h}_{N_{t}}] \in \mathbb{C}^{N_{r}\times N_{t}}$} is the channel matrix with unit variance. Each column vector of $\mathbf{H}$ is represented by $\mathbf{h_{i}}$. In this work, the channel matrix $\mathbf{H}$ is perfectly known to the receiver and flat Rayleigh fading. The total transmitted energy is set to unity and equally shared by transmit antennas, $\mathrm{tr}\left(\mathrm{E}[\mathbf{x} {\mathbf{x}}^{\text{H}}] \right /N_{t})=1$. The transmitted vector $\mathbf{x}$ can be written as the sum of the base layer vector and enhancement layer vector,  $\mathbf{x}=\mathbf{x}_{b}+\mathbf{x}_{e}$. The received signal vector in (\ref{eq:basicMIMO}) is rewritten as          
\begin{align}
\mathbf{y}=\frac{1}{\sqrt{N_{t}}}\mathbf{H}\mathbf{x}_{b} + 
\frac{1}{\sqrt{N_{t}}}\mathbf{H}\mathbf{x}_{e} + \mathbf{n},
\label{eq:basicMIMO2}
\end{align} 
where $\mathbf{x}_{b}=[x_{b_{1}}, ... , x_{b_{N_{t}}}]^{\text{T}}\in \mathbb{C}^{N_{t}\times1}$ is the transmitted base layer vector with 
the property \mbox{$\mathrm{E}[\mathbf{x}_{b}{\mathbf{x}_{b}}^{\text{H}}] = E_{x_{b}}\mathbf{I}$} and $\mathbf{x}_{e} = [x_{e_{1}}, ... ,x_{e_{N_{t}}}]^{\text{T}}\in \mathbb{C}^{N_{t}\times1}$ is the transmitted enhancement layer vector with $\mathrm{E}[\mathbf{x}_{e}{\mathbf{x}_{e}}^{\text{H}}] = E_{x_{e}}\mathbf{I}$. The total transmitted symbol energy equals 1, i.e., $E_{x_{b}} + E_{x_{e}} = 1$. 

\section{Proposed Receiver Structure}
\label{sec:ProposedReciver}
We propose a novel MIMO transceiver structure with a low computational receiver complexity, which is shown in \mbox{Fig. \ref{fig:Structure}}. In the transmitter part, hierarchical modulation is used to enable the sequential detection. In the receiver, MMSE-ML detection order is used. First, the base layer is detected with MMSE receiver which is followed by cancellation of the base layer based on hard decision. Finally, the enhancement layer is detected with ML.  

\subsection{Detection Order}
\label{sec:WhyOrder}
\begin{figure}[]
\centering
\pgfplotsset{/pgf/number format/use comma,compat=newest, width=1cm, height=1cm, every axis/.append style={font=\scriptsize}}	
\definecolor{mycolor1}{rgb}{0.00000,0.74902,0.74902}
\definecolor{mycolor2}{rgb}{0.00000,0.49804,0.00000}
\begin{tikzpicture}

\begin{axis}[
width=7.5cm,
height=4.79cm, 
scale only axis,
xmin=0,
xmax=35,
xlabel={$E_{b}/N_{0}$ (dB)},
xmajorgrids,
ymode=log,
ymin=0.0001,
ymax=1,
yminorticks=true,
ylabel={BER},
ymajorgrids,
yminorgrids,
legend style={draw=black,fill=white,legend cell align=left,at={(1,1)}} ]

\addplot [color=blue,solid,line width=1.0pt,mark=o,mark options={solid}]
table[row sep=crcr]{
0 0.0682301899148657\\
5	0.0251250602700096\\
10	0.0128251482142491\\
15	0.00907963721389567\\
20	0.00739030500703135\\
25	0.00746195165622202\\
30	0.00730630409643621\\
35	0.00735449828821516\\ };
\addlegendentry{Base, 16-HQAM, $d$=4};

\addplot [color=mycolor1,dashed,line width=1.0pt,mark=o,mark options={solid}]
table[row sep=crcr]{
0	0.316210980135342\\
5	0.197705721632915\\
10	0.0872438303895624\\
15	0.0260725281747415\\
20	0.0105566289691684\\
25	0.00919128618322889\\
30	0.00863936877270444\\
35	0.00874863233685102\\ };
\addlegendentry{Enh., 16-HQAM, $d$=4};

\addplot [color=red,solid,line width=1.0pt,mark=star,mark options={solid}]
table[row sep=crcr]{
0	0.0450780517701348\\
5	0.0106340015983948\\
10	0.00238271541405547\\
15	0.000851944761336207\\
20	0.000541668507971333\\
25	0.000450718447637225\\
30	0.000432918534956235\\
35	0.000429316500314572\\ };
\addlegendentry{Base, 16-HQAM, $d$=8};

\addplot [color=mycolor2,dashed,line width=1.0pt,mark=star,mark options={solid}]
table[row sep=crcr]{
0	0.38880777273055\\
5	0.306409308099101\\
10	0.201829885412378\\
15	0.0900801463562899\\
20	0.0222439351756778\\
25	0.00361457371914724\\
30	0.000869213496842435\\
35	0.000565610859728507\\ };
\addlegendentry{Enh., 16-HQAM, $d$=8};

\end{axis}
\end{tikzpicture}
\caption{BER for uncoded -- block fading channel model is not utilized -- \mbox{$2 \times 2$ MIMO} system, where ML receivers operate in both layers.} 
\label{fig:MLML} 
\end{figure}
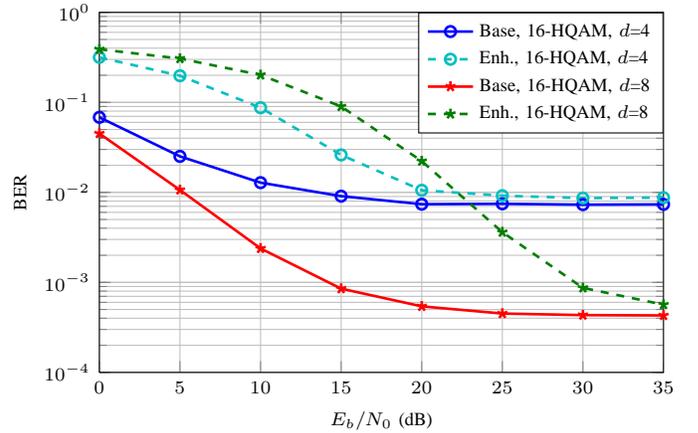

Let us first focus on the performance of the receiver, where ML receivers operate at both layers as in \cite{rupp}. As seen from \mbox{Fig. \ref{fig:MLML}}, error floor arises due to the interference of the enhancement layer to the base layer. ML detection in the first stage does not account for the interference caused by the enhancement layer. However, a better performance is attained with the proposed MMSE-ML detection order. This is due to the fact that, as opposed to the ML detection, interference from the enhancement layer is minimized by the first-stage MMSE receiver which helps the proposed structure operate well. The system has a lower overall interference after MMSE based base layer detection. After base layer detection, the base layer ideally has no interference influencing the enhancement layer so that enhancement layer can be smoothly detected with ML.  

\subsection{Why Block Fading Channel?}
\label{sec:WhyBlock}
Under a spatial multiplexing transmitter structure, MMSE receiver always has diversity $N_{r}-N_{t}+1$ \cite{hedayat}, so that spatial diversity is not obtained for the case $N_{t}=N_{r}$, which is our focus here. Diversity of MMSE receiver is limited to unity unless another diversity domain (e.g. frequency diversity, time diversity) is present in the system. With the help of multiple independently fading blocks, some level of diversity can be attained along with coding. Hence, a block fading channel model \cite{knopp} is utilized to achieve diversity with an MMSE receiver in this work. A single frame is partitioned into $F$ blocks that are transmitted over different independently fading carrier frequencies, therefore, frequency diversity is obtained. Channels are constant within a block and change independently between blocks, which is a valid assumption for various communication systems, such as slow frequency hopping systems, multi-carrier schemes, etc.

\subsection{Log Likelihood Ratio Computation}
\label{sec:LLR}
The MMSE equalization filter is found by minimizing the mean square error as
\begin{align}
\mathbf{W}^{\text{H}} = 
\frac{E_{x_{b}}}{\sqrt{N_{t}}}\mathbf{H}^{\text{H}}\left(\frac{1}{N_{t}}\mathbf{H}\mathbf{H}^{\text{H}}+N_{0}\mathbf{I}\right)^{-1},
\label{eq:MMSE}
\end{align}
and the Hermitian transpose of the MMSE filter is written as 
\begin{align}
\mathbf{W}=[\mathbf{g}_{1},\mathbf{g}_{2}, ... ,\mathbf{g}_{N_{t}}],
\label{eq:W}
\end{align}
where $\mathbf{g}_{i}$ is the filter vector which produces the $i$-th output of the MMSE filter. The filter vector is represented as
\begin{align}
\mathbf{g}_{i} = 
\frac{E_{x_{b}}}{\sqrt{N_{t}}}\left(\frac{1}{N_{t}}\mathbf{H}\mathbf{H}^{\text{H}}+N_{0}\mathbf{I}\right)^{-1}{\mathbf{h}}_{i}.
\label{eq:filterVec}
\end{align}
Passing the received vector through the filter vectors yields
\begin{align}
{z_{b}}_{i}={\mathbf{g}_{i}}^{\text{H}}\mathbf{y}=\beta_{i}{x_{b}}_{i}+\eta_{i},
\label{eq:vecPass}
\end{align}
where $\beta_{i}=\frac{1}{\sqrt{N_{t}}}{\mathbf{g}_{i}}^{\text{H}}\mathbf{h}_{i}$ and $\eta_{i}$ is the interference-plus-noise term modeled as a complex Gaussian random variable given by
\begin{align}
\eta_{i}=\frac{1}{\sqrt{N_{t}}}{\mathbf{g}_{i}}^{\text{H}}\mathbf{h}_{i}{x_{e}}_{i} + \sum_{k\neq i}^{}\frac{1}{\sqrt{N_{t}}}{\mathbf{g}_{i}}^{\text{H}}\mathbf{h}_{k}x_{k} + {\mathbf{g}_{i}}^{\text{H}}\mathbf{n}.
\label{eq:interference}
\end{align}
At the $i$-th MMSE receiver output, the log likelihood ratio (LLR) corresponding the $j$-th base layer bit is calculated as
\begin{align}
\Lambda_{b}^{(j,i)}&= \ln \frac 
{\Pr(b^{j}=1\vert z_{b_{i}}, \beta_{i}, \eta_{i})} 
{\Pr(b^{j}=0\vert z_{b_{i}}, \beta_{i}, \eta_{i})}  \nonumber\\
&= \ln \frac 
{\sum_{\tilde{x}_{b}\in B_{1}^{j}}^{} \Pr(z_{b_{i}} \mid 
	\tilde{x}_{b},\beta_{i},\eta_{i})} 
{\sum_{\tilde{x}_{b}\in B_{0}^{j}}^{} \Pr(z_{b_{i}} \mid 
	\tilde{x}_{b},\beta_{i},\eta_{i})} \nonumber\\
&= \ln \frac 
{\sum_{\tilde{x}_{b}\in B_{1}^{j}}^{} \exp \left(-\frac{{\vert 
			\beta_{i}({x_{b}}_{i}-\tilde{x}_{b}) 
			\vert}^{2}}{{{\sigma}_{\eta_{i}}}^{2}}\right) 
}    
{\sum_{\tilde{x}_{b}\in B_{0}^{j}}^{} \exp\left(-\frac{
		{\vert 
			\beta_{i}({x_{b}}_{i}-\tilde{x}_{b}) 
			\vert}^{2}   }{{{\sigma}_{\eta_{i}}}^{2}}\right)}~,  
\label{eq:LLRmmse}  
\end{align}
where $B_{1}^{j}$ and $B_{0}^{j}$ are the the subsets of the base layer constellation points with $j$-th bit equal to $1$ and $0$, respectively \cite{kim}. After LLRs for the base layer are determined and passed into the base layer decoder, hard decisions are made for the base layer symbols and then cancellation of the base layer is performed. Finally LLRs based on the ML receiver for the enhancement layer are calculated \cite{butler} and then given to the decoder.  

\section{Error Rate Performance Results}
\label{sec:SimResults}
In this section, we inspect the proposed receiver's performance with the help of the following setup. WiMAX LDPC codes \cite{wimax} -- defined in the IEEE 802.16e standard -- are used. The frame length is fixed to 2304 bits, which is the largest code length of WiMAX LDPC codes. To do that, a sequence of 2304 bits are generated at the output of the encoder for standard 16-QAM and 1152 bits are generated at the output of each layer's encoder for 16-HQAM. The maximum number of iterations are set to 50 in the LDPC decoder. Each frame is transmitted over $F=8$ blocks that have independent and identically distributed fading. The code rates of the base layer and the enhancement layer are expressed as $R_{b}$ and $R_{e}$, respectively. The overall code rate is defined as $R=(R_{b}+R_{e})/2$. A high spectral efficiency is desired in the system, therefore an overall system code rate of $R=3/4$ is chosen. Frame error rate (FER) performances of the proposed receiver are examined and compared to that of MMSE and ML receivers for $2 \times 2$ and $4 \times 4$ MIMO systems.           

\begin{figure}[]
\centering
\pgfplotsset{/pgf/number format/use comma,compat=newest, width=1cm, height=1cm, every axis/.append style={font=\scriptsize} }		
\definecolor{mycolor1}{rgb}{0.74902,0.00000,0.74902}
\definecolor{mycolor2}{rgb}{0.00000,0.49804,0.00000}
\begin{tikzpicture}
	
\begin{axis}[
width=7.5cm,
height=5.35cm,
scale only axis,
xmin=4,
xmax=16,
xlabel={$E_{b}/N_{0}$ (dB)},
xmajorgrids,
ymode=log,
ymin=1e-05,
ymax=1,
yminorticks=true,
ylabel={FER},
ymajorgrids,
yminorgrids,
legend style={draw=black,fill=white,legend cell align=left,at={(1,1)}} ]
	
\addplot [color=red,solid,line width=1.0pt,mark=o,mark options={solid}]
table[row sep=crcr]{
0	1\\
1	0.985294117647059\\
2	0.881578947368421\\
3	0.688356164383562\\
4	0.446666666666667\\
5	0.283098591549296\\
6	0.118374558303887\\
7	0.0518709677419355\\
8	0.0201888308557654\\
9	0.00661380013819881\\
10	0.0019742657892152\\
11	0.000676134796385874\\
12	0.000207342610453781\\
13	5.65651994731356e-05\\
14	1.7458260209549e-05\\ };
\addlegendentry{Base, $d$=2, $R_{b}$=2/3};
	
\addplot [color=blue,dashed,line width=1.0pt,mark options={solid}]
table[row sep=crcr]{
0	1\\
1	1\\
2	1\\
3	1\\
4	1\\
5	0.995774647887324\\
6	0.962308598351001\\
7	0.79741935483871\\
8	0.506629168340699\\
9	0.235036688493304\\
10	0.0761025439544249\\
11	0.0197222801552755\\
12	0.00421906107838789\\
13	0.000812114649578589\\
14	0.000167100490577112\\ };
\addlegendentry{Enh., $d$=2, $R_{e}$=5/6};
	
\addplot [color=mycolor1,solid,line width=1.0pt,mark=star,mark options={solid}]
table[row sep=crcr]{
0	1\\
1	1\\
2	1\\
3	0.990147783251232\\
4	0.934883720930233\\
5	0.770114942528736\\
6	0.603603603603604\\
7	0.448660714285714\\
8	0.229714285714286\\
9	0.120143454871488\\
10	0.0635071090047393\\
11	0.0282819755170958\\
12	0.012348712907784\\
13	0.00480481916190567\\
14	0.00193796581081211\\
15	0.000911374497837186\\
16	0.000390002735840087\\ };
\addlegendentry{Base, $d$=2, $R_{b}$=3/4};
	
\addplot [color=mycolor2,dashed,line width=1.0pt,mark=square,mark options={solid}]
table[row sep=crcr]{
0	1\\
1	1\\
2	1\\
3	1\\
4	0.990697674418605\\
5	0.904214559386973\\
6	0.744744744744745\\
7	0.511160714285714\\
8	0.264\\
9	0.128511655708308\\
10	0.0647709320695103\\
11	0.0285633882088082\\
12	0.012348712907784\\
13	0.00480481916190567\\
14	0.00193796581081211\\
15	0.000911374497837186\\
16	0.000390002735840087\\ };
\addlegendentry{Enh., $d$=2, $R_{e}$=3/4};
	
\end{axis}
\end{tikzpicture}
\caption{$2 \times 2$ MIMO system, FER performances of the proposed receiver (16-HQAM) with fixed $d=2$ for different code rates.} 
\label{fig:codedMIMO2x2_fixed_d} 
\end{figure}
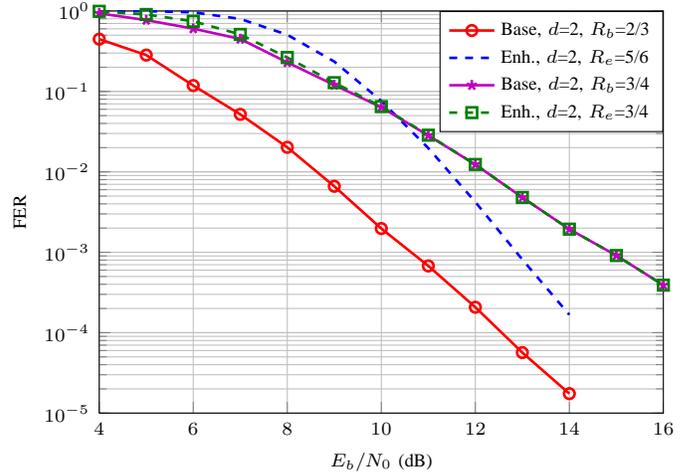

Before inspecting the proposed structure's performance with optimized design parameters, we like to stress out the importance of how to choose code rates of layers. In Fig. \ref{fig:codedMIMO2x2_fixed_d}, the constellation ratio $d$ of 16-HQAM is fixed to 2 and the effect of code rates are examined. The performance gap between layers is higher when $R_{b}=2/3$ and $R_{e}=5/6$. Using the same code rate in both layers ($R_{b}=R_{e}=3/4$) brings the performances of layers closer yet at the cost of worse overall FER. The effect of code rates on the performance can be justified considering an upper bound to diversity based on the Singleton bound for SISO systems ($\text{diversity} \leq \lfloor F(1-R)\rfloor+1$) \cite{malkamaki}. Increasing $R_{b}$ from $2/3$ to $3/4$ results in poor diversity order in the base layer. Decreasing $R_{e}$ from $5/6$ to $3/4$ may be expected to lead to a higher diversity order in the enhancement layer, which is immaterial since the enhancement layer performance is limited by that of the base layer. As follows from the preceding arguments, one has to be careful about rate allocation for the layers. Base layer's code rate should be usually low so that sufficient diversity to reach desired error performance can be created based on frequency diversity since spatial diversity is not possible with MMSE receivers in our setting. Higher code rates at the enhancement layer keeps the overall code rate constant. 

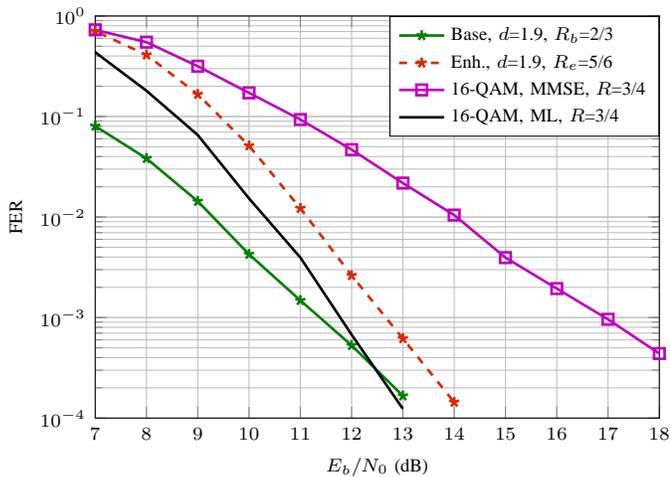
\begin{figure}[]
\centering
\pgfplotsset{/pgf/number format/use comma,compat=newest, width=1cm, height=1cm, every axis/.append style={font=\scriptsize}}		
\definecolor{mycolor1}{rgb}{0.00000,0.49804,0.00000}
\definecolor{mycolor2}{rgb}{0.84706,0.16078,0.00000}
\definecolor{mycolor3}{rgb}{0.74902,0.00000,0.74902}
\begin{tikzpicture}
	
\begin{axis}[
width=7.5cm,
height=5.35cm,
scale only axis,
xmin=7,
xmax=18,
xtick={ 0,  1,  2,  3,  4,  5,  6,  7,  8,  9, 10, 11, 12, 13, 14, 15, 16, 17, 18},
xlabel={$E_{b}/N_{0}$ (dB)},
xmajorgrids,
ymode=log,
ymin=0.0001,
ymax=1,
yminorticks=true,
ylabel={FER},
ymajorgrids,
yminorgrids,
legend style={draw=black,fill=white,legend cell align=left,at={(1,1)}} ]
	
\addplot [color=mycolor1,solid,line width=1.0pt,mark=star,mark options={solid}]
table[row sep=crcr]{
0	1\\
1	0.995049504950495\\
2	0.939252336448598\\
3	0.788235294117647\\
4	0.654723127035831\\
5	0.38212927756654\\
6	0.209593326381648\\
7	0.0803357314148681\\
8	0.0382930081920366\\
9	0.0143653516295026\\
10	0.00426380433168579\\
11	0.00148162344650676\\
12	0.000528712230613227\\
13	0.000166524030411762\\ };
\addlegendentry{Base, $d$=1.9, $R_{b}$=2/3};
	
\addplot [color=mycolor2,dashed,line width=1.0pt,mark=star,mark options={solid}]
table[row sep=crcr]{
0	1\\
1	1\\
2	1\\
3	1\\
4	1\\
5	0.986692015209126\\
6	0.940563086548488\\
7	0.705835331734612\\
8	0.410554391312631\\
9	0.166237850200114\\
10	0.0512080778939776\\
11	0.0122289218793767\\
12	0.00262251788020591\\
13	0.000618044411378976\\
14	0.000144003748396078\\ };
\addlegendentry{Enh., $d$=1.9, $R_{e}$=5/6};
	
\addplot [color=mycolor3,solid,line width=1.0pt,mark=square,mark options={solid}]
table[row sep=crcr]{
0	1\\
1	1\\
2	1\\
3	1\\
4	1\\
5	0.966346153846154\\
6	0.851694915254237\\
7	0.728260869565217\\
8	0.549180327868853\\
9	0.316535433070866\\
10	0.171941830624465\\
11	0.0934014869888476\\
12	0.0468531468531469\\
13	0.0217838950904953\\	
14	0.0104665694646949\\
15	0.00394930739758326\\
16	0.00194458418791842\\
17	0.00096120279466126\\
18	0.000438608933220153\\ };
\addlegendentry{16-QAM, MMSE, $R$=3/4};
	
\addplot [color=black,solid,line width=1.0pt]
table[row sep=crcr]{
0	1\\
1	1\\
2	1\\
3	1\\
4	0.990147783251232\\
5	0.881578947368421\\
6	0.636075949367089\\
7	0.436008676789588\\
8	0.180430879712747\\
9	0.0654723127035831\\
10	0.0153201219512195\\
11	0.00395895294557917\\
12	0.000675968804544125\\
13	0.00012372154404487\\ };
\addlegendentry{16-QAM, ML, $R$=3/4};
	
\end{axis}
\end{tikzpicture}
\caption{$2 \times 2$ MIMO system, FER performances of ML, MMSE and the proposed receiver (16-HQAM).} 
\label{fig:codedMIMO2x2} 
\end{figure}

\begin{figure}[]
\centering
\pgfplotsset{/pgf/number format/use comma,compat=newest, width=1cm, height=1cm, every axis/.append style={font=\scriptsize}}			
\definecolor{mycolor1}{rgb}{0.00000,0.49804,0.00000}
\begin{tikzpicture}
	
\begin{axis}[
width=7.5cm,
height=5.35cm,
scale only axis,
xmin=4,
xmax=14,
xtick={ 0,  1,  2,  3,  4,  5,  6,  7,  8,  9, 10, 11, 12, 13, 14, 15, 16, 17, 18},
xlabel={$E_{b}/N_{0}$ (dB)},
xmajorgrids,
ymode=log,
ymin=0.0001,
ymax=1,
yminorticks=true,
ylabel={FER},
ymajorgrids,
yminorgrids,
legend style={draw=black,fill=white,legend cell align=left,at={(1,1)}} ]

\addplot [color=red,solid,line width=1.0pt]
table[row sep=crcr]{
0	0.95260663507109\\
1	0.794466403162055\\
2	0.512755102040816\\
3	0.248454882571075\\
4	0.0673141326188881\\
5	0.0212946286682911\\
6	0.00455431186840078\\
7	0.000882639663455205\\
8	0.000176793911932794\\ };
\addlegendentry{Base, $d$=2, $R_{b}$=2/3};
	
\addplot [color=blue,dashed,line width=1.0pt]
table[row sep=crcr]{
0	1\\
1	1\\
2	1\\
3	0.980222496909765\\
4	0.699263228399196\\
5	0.172475897870537\\
6	0.0132324285131645\\
7	0.00111976673721929\\
8	0.000184739705727527\\ };
\addlegendentry{Enh., $d$=2, $R_{e}$=5/6};
	
\addplot [color=mycolor1,solid,line width=1.0pt,mark=o,mark options={solid}]
table[row sep=crcr]{
0	1\\
1	1\\
2	1\\
3	1\\
4	1\\
5	0.926267281105991\\
6	0.773076923076923\\
7	0.511450381679389\\
8	0.272357723577236\\
9	0.113559322033898\\
10	0.0456506927095162\\
11	0.0160146601864393\\
12	0.00543360726643599\\
13	0.00202958550007573\\
14	0.000584227598795503\\ };
\addlegendentry{16-QAM, MMSE, $R$=3/4};
	
\end{axis}
\end{tikzpicture}	
\caption{$4 \times 4$ MIMO system, FER performances of MMSE and the proposed receiver (16-HQAM).}
\label{fig:codedMIMO4x4} 
\end{figure}
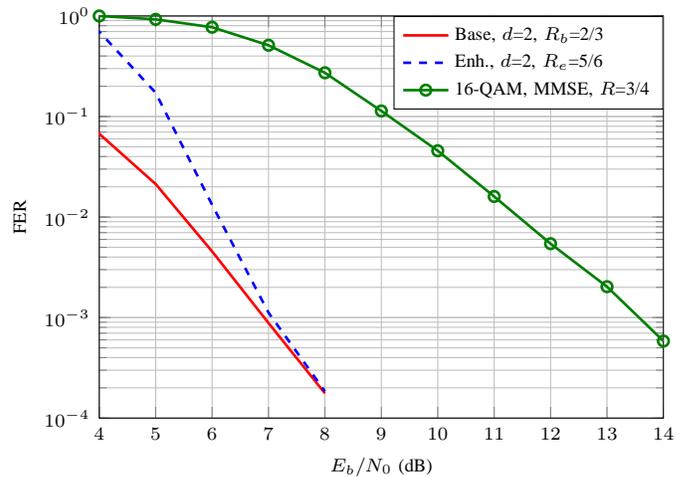

After carefully choosing the code rates of the layers, how to choose the constellation $d$ is also important. The ratio $d$ needs to be adjusted such that sufficient protection is supplied to the base layer by choosing $d$ large enough, yet also to the enhancement layer by choosing $d$ small enough. Here we show the error performances with an optimized constellation ratio $d$.  Fig. \ref{fig:codedMIMO2x2} shows a $2 \times 2$ system's FER performance of ML with $R=3/4$, of MMSE with $R=3/4$ and of the proposed receiver structure with $d=1.9$, $R_{b}=2/3$ and $R_{e}=5/6$. For $\mathrm{FER} = 10^{-3}$, the average performance of the proposed structure is around 0.5 dB off from the ML detector. Moreover, the proposed structure shows around 4.5 dB better performance than the MMSE detector.  Fig. \ref{fig:codedMIMO4x4} shows a $4 \times 4$ system's FER performance of MMSE and the proposed receiver structures with $d=2$, $R_{b}=2/3$ and $R_{e}=5/6$. For $\mathrm{FER} = 10^{-3}$, the proposed structure outperforms MMSE receiver by around 6.5 dB. In Fig. \ref{fig:codedMIMO4x4}, ML detector performance is not depicted since statistically significant numerical results were not available due to the very high running time of the ML detector. Please notice that the overall performance of hierarchical modulation which is simply the arithmetic mean of error rate of layers, is not plotted in the figures, for the sake of clear illustration.

Due to the page constraint, we would like to summarize the complexity of the proposed structure without details. Since 16-HQAM detection is composed of two 4-QAM detection stages and complexity of the ML receiver is higher than that of MMSE, the overall complexity of the proposed receiver is dominated by ML detection complexity -- especially for a high number of antennas -- and can be approximately written as $\mathcal{O}(N_{t}{4}^{N_{t}})$ per one MIMO symbol vector. Although not presented here, similar performance and computational complexity results are observed for higher constellations such as 64-QAM where three layers are employed. In the proposed receiver for 64-QAM, the first two layers (the base layer and the enhancement layer with higher minimum distances) are detected with MMSE receivers and the enhancement layer is detected with ML in the last step.

\section{Conclusion}
\label{sec:Conclusion}
In this work, a new receiver structure with low computational complexity is proposed for MIMO systems. The idea makes use of hierarchical modulation so 
that processing is performed sequentially for each layer. The proposed scheme provides a performance between that of ML receiver and MMSE receiver at a 
significant lower complexity compared to ML. MMSE detection in the base layer reduces interference power and enhances performance. ML detection at the enhancement layer with 4-QAM enjoys good performance with low computational complexity. Performance of each layer is enhanced through coding over multiple fading blocks. With carefully chosen constellation ratio and coding rates, performance quite close to that of ML receiver can be achieved. Moreover, receiver computational complexity drops from $\mathcal{O}(N_t4^{P N_{t}})$ to $\mathcal{O}(N_t 4^{N_{t}})$ with $4^{P}$-HQAM modulation, where $P$ is the number of layers. This is a significant complexity advantage, especially for MIMO systems with a high number of antennas. 
  
\bibliographystyle{IEEEtran}
\bibliography{YigitCommRef}
\end{document}